\documentclass[prd,a4paper,nofootinbib,
showpacs,
twocolumn
]{revtex4}
\usepackage{graphicx}
\def\cc{WZ}

\def\Tr{\mbox{Tr}\,}

\def\di{\mbox{d}}

\def\hbar{\hspace{0pt}\raisebox{1pt}{$-$} \hspace{-7pt} h}

\def\Z{{\bf Z}}
\def\5{\overline 5}

\newcommand{\be}{\begin{equation}}
\newcommand{\ee}{\end{equation}}
\newcommand{\bea}{\begin{eqnarray}}
\newcommand{\eea}{\end{eqnarray}}
\newcommand{\nn}{\nonumber}

\begin{document}
\title[Anomaly Cancellation]{Cancellation of Global Anomalies in Spontaneously
Broken Gauge Theories}
\date{\today}
\author{M.~Fabbrichesi}
\author{R.~Percacci}
\author{M.~Piai}
\author{M.~Serone}

\affiliation{INFN, Sezione di Trieste and\\
Scuola Internazionale Superiore di Studi Avanzati\\
via Beirut 4, I-34014 Trieste, Italy.}
\begin{abstract}

\noindent We discuss the generalization to global gauge anomalies
of the familiar procedure for the cancellation of local gauge anomalies
in effective theories of spontaneously broken symmetries.
We illustrate this mechanism in a recently proposed
six-dimensional extension of the standard model.

\end{abstract}
\pacs{11.10.Kk,11.15.Ex,11.30.-j}
\maketitle
%
\vskip1.5em
\section{Introduction}

It is always possible to cancel gauge anomalies in effective theories
by introducing non-renormalizable operators. These operators contain boson fields
that transform non-linearly under the corresponding gauge symmetries.
These fields can either be introduced as
additional degrees of freedom---as in the Green-Schwarz (GS)~\cite{GS} and
other mechanisms~\cite{archaeology}---or be the would-be Goldstone bosons of the spontaneous
breaking of the gauge symmetry~\cite{Preskill}.

After briefly reviewing the  literature on local (perturbative) gauge
anomalies, we discuss in some detail the global
(non-perturbative) case~\cite{Wittensu2} in $d$-dimensional effective theories of a gauge group $G$
spontaneously broken to $H$. We show how to construct local operators that cancel both local and global anomalies, using the would-be Goldstone bosons of the spontaneously broken theory without introducing additional degrees of freedom.
 This can be done
provided that:~\footnote{A possible limitation of this procedure arises whenever we wish to preserve  further symmetries of the original action that are explicitly broken  by the introduction of these operators.}
\begin{itemize}
\item[1.] the coset space $G/H$ is reductive;
\item[2.] the fermion representations are free of \textit{local} anomalies when restricted to
the group $H$;
\item[3.] the fermion representations are free of  \textit{global} anomalies when restricted
to the group $H$;
\item[4.] $G$ can be embedded in a  group $K$ such that its homotopy group $\pi_d(K)=0$ and
the fermion representations can be extended to $K$ without generating further anomalies of $G$.
\end{itemize}
Most of  these conditions can be derived from the literature.
In particular, conditions 1 and 2 have been  derived for local gauge anomalies~\cite{Wu,AGG,CHZ},
while conditions 3 and 4 are closely related to the results of~\cite{WittenGlobal,w,en}.
In this paper---generalizing the idea of computing global anomalies as local anomalies of
a larger group $K$~\cite{en}---we argue that the operators constructed in \cite{Wu,AGG,CHZ} to
cancel local anomalies of $G$ can be properly defined globally and used to cancel
global anomalies of $G$, provided the above four conditions are satisfied.

This result is of interest in model building in so far as the fermion content of an effective
theory is only restricted by the cancellation of the anomalies of the unbroken group $H$.
As an example we show how the standard model in $d=6$~\cite{DP} can be made (local and global)
anomaly free.

\section{Cancellation of Local Gauge Anomalies}

In purely bosonic low-energy effective actions, anomalies are accounted for
by the Wess-Zumino (\cc) term~\cite{WZ}.
For a global symmetry group $G$ spontaneously broken to
an anomaly-free subgroup $H$,  the \cc\ coupling is a non-invariant function of
the corresponding Goldstone bosons.
The same term can be generalized to reproduce also gauge anomalies when part or all
of the group $G$ is gauged~\cite{WittenGlobal}.
 From an effective field theory point of view, the \cc\ term can be seen
as the remnant of massive fermions that have been integrated out
in a microscopic theory~\cite{hokerfari}. This is in agreement
with the 't Hooft anomaly matching condition~\cite{tHooft}:
anomalies, being long-distance effects, have to match in going
from the fundamental to the effective  theory.

The \cc\ term can also be viewed from a different perspective:
an anomalous theory can be rendered gauge invariant by coupling it
to $G$-valued scalar fields and adding the \cc\ term to the action  \cite{archaeology}.
Thus, instead of \textit{reproducing} the anomalies of an underlying theory,
one can use the \cc\ term to \textit{cancel} them.
This construction is similar to the GS mechanism~\cite{GS},
which also requires the presence of additional  degrees of freedom
(but only works for reducible anomalies).

If $\Gamma(A)$ is the effective action for the gauge field $A$ obtained by integrating out fermions,
the \cc\ term is defined in general by
\be
\Gamma_{\cc\ }(A,g)=\Gamma(A^g)-\Gamma(A) \, ,
\label{wzdef}
\ee
where $A^g = g^{-1} A g + g^{-1} dg$ is the gauge transformed
connection. From the definition, it satisfies the following (cocycle) condition
\be
\Gamma_{\cc\ }(A^g,U^g)-\Gamma_{\cc\ }(A,U)=-\Gamma_{\cc\ }(A,g) \, ,
\label{cocycle}
\ee
where $U$ are the $G$-valued scalar fields and $U^g=g^{-1} U$.
If the low-energy theory contains both fermions and scalars,
and the scalar action contains the \cc\ term, then the effective action
$\tilde\Gamma(A,U)=\Gamma(A)+\Gamma_{\cc\ }(A,U)$
is seen to be gauge invariant  by means of (\ref{wzdef}) and (\ref{cocycle}).
It can be shown that the addition of the \cc\ term cures also the problems
arising in the canonical formulation of the theory~\cite{perca2}, so that the quantum theory,
aside from renormalizability issues, is well defined.

Although the fermionic effective action
is nonlocal, the \cc\ action is local.
The standard way of deriving an explicit expression for the \cc\ action
is through the dimensional descent.
Assume that the boundary conditions on the fields are such that
spacetime can be compactified to a $d$-dimensional sphere (we assume $d$ to be even).
We can think of this sphere as the boundary of a $d+1$-dimensional ball $B$.
Assume for simplicity that the gauge fields are topologically trivial (no instantons)
and that the scalar fields are homotopically trivial.
(This will automatically be the case if the homotopy group $\pi_d(G)=0$,
but we need not assume this).
Then the fields can be extended to a gauge field $\hat A$ and a
scalar field $\hat U$ defined on all of $B$.
Let $\Omega(\hat A)$ be the Chern-Simons functional, the integral of the
 $(d+1)$-form $\omega(\hat A)$ on $B$. Its gauge variation under a gauge
transformation $g$ is the \cc\ action:
\be
\Gamma_{\cc\ }(A,g)=\Omega(\hat A^{\hat g})-\Omega(\hat A) \,.
\label{wzexpl}
\ee
Even though the right-hand side of (\ref{wzexpl}) is written in terms of the fields on $B$, it is
the integral of an exact form and therefore can be rewritten as an integral
on spacetime, depending only on the boundary values of the fields $\hat A$ and $\hat U$.
This gives an explicit, local formula for the \cc\ term that cancels
local anomalies.

A variation of this anomaly-cancellation scheme occurs when the scalar fields are not
introduced by hand but are already present in the theory.
This can happen in the Higgs phase of a gauge theory~\cite{Preskill}. The would-be Goldstone bosons---the angular
components of the Higgs field---are the scalar fields that can be used to write the \cc\ term.
 This idea had been already put forward in \cite{GJ}
where the interplay between renormalizability and anomaly cancellation in
spontaneously broken gauge theories was first discussed in an  abelian Higgs
model coupled to
a chiral fermion in four dimensions. This perspective---that spontaneously broken
effective gauge  theories with anomalous fermion content can be consistently
quantized---has been exploited in~\cite{Preskill}
where the approach is extended to
non-abelian gauge symmetries.

Since there is usually a nontrivial unbroken group $H$, one has
to generalize the \cc\ term to the case of $G/H$-valued, rather than $G$-valued,
scalars $\varphi$.
Let us review more explicitly how this construction works, following closely the analysis
in \cite{Wu,AGG,CHZ}  from the perspective of \cite{Preskill}.
Consider a given $d$-dimensional action with a local gauge symmetry $G$ spontaneously
broken to a subgroup $H$. We denote by $T^A$ the whole set of generators of the Lie algebra
$\cal G$ of $G$, whereas $T^i$ and $T^\alpha$ are respectively the generators of $\cal H$ and of the
coset $G/H$. We assume that $G/H$ is a reductive space, i.e.\  that
$[T^i, T^\alpha] = f_{i \alpha \beta} T^\beta$. Assume also that the fermion content of the
corresponding action gives rise to the following anomaly:
\bea
\delta_\alpha \Gamma(A) & = & \int d^dx \,  {\cal A}_\alpha [A(x)] \, ,\nn  \\
\delta_i \Gamma(A) & = & 0,
\label{anomalia}
\eea
where ${\cal A}_\alpha$ denotes the usual one-loop gauge anomaly. In other words, eq.(\ref{anomalia})
implies that any potentially anomalous fermionic one-loop amplitude vanishes as soon as one of the
external gauge fields belongs to $H$.

To write a \cc\ term, we now assume that there exists a unitary gauge,
i.e. that there is a globally defined gauge transformation $U(x)$ that
transforms the $G/H$-valued field $\varphi(x)$ to a constant.~\footnote{This is always true whenever the gauge field $A$ and its components in $H$ are topologically trivial. More general cases can be
however similarly analyzed.}
This is a purely topological restriction on $\varphi$; it is less
restrictive than the assumption of being homotopic to a constant.

This gauge transformation $U$ is not unique:
two $G$-valued maps $U$ and $U'$ correspond to the same $\varphi$
if and  only if they differ by a right-$H$ transformation: $U'(x)=U(x)h(x)$.
Thus we can use $U$ as a dynamical variable instead of $\varphi$,
but in doing so, we introduce an additional $H$ gauge freedom.
Having reformulated the theory in terms of $U$, we can add to the action
a \cc\ term $\Gamma_{WZ}(A,U)$. Because of (\ref{anomalia}),
$\Gamma_{WZ}(A,Uh)=\Gamma_{WZ}(A,U)$, so it depends only on the
coset $\varphi(x)=U(x)H$, i.e. on the would-be Goldstone fields.
In this case we will therefore write
$\Gamma_{WZ}(A,\varphi)=\Gamma_{WZ}(A,U)$.
The addition of this term to the action cancels the anomaly
exactly as in the previous case.~\footnote{One often encounters the notation $U(x)=e^{\xi^\alpha(x)
T_\alpha}$; this amounts to choosing a specific coordinate system
on the coset space and can generally be valid only locally.
In this notation
$\Gamma_{WZ}(A,\xi) =  - \int_0^1 \!dt \int \! d^d x \; \xi^\alpha (x)
\, {\cal A}_\alpha [A_t]$
where $A_t = e^{-t\xi} A e^{t\xi} + e^{-t\xi} d e^{t\xi}$.
In the simple case of $G=U(1)$ and $H=\{1\}$ one can easily check
that this reduces to the form $\int \xi F \tilde F$ \cite{GJ},
as also discussed in~\cite{DSW} in the context of the GS mechanism in $d=4$.}

This procedure can be used also when the second condition in
eq.~(\ref{anomalia}) is replaced by the weaker condition $\delta_i \Gamma(A_H)=0$
where $A_H$ denotes the component of the gauge field
in the subalgebra of $H$,
i.e. when the fermion representations, restricted to the subgroup
$H$, are free of local anomalies.
As shown in \cite{Wu,AGG,CHZ}, one can add to the action
a local functional $B_d(A_H,A)$ such that
the second relation in (\ref{anomalia}) is satisfied
(see \cite{CHZ} and eq.~(\ref{bd}) below for an explicit expression of $B_d(A_H,A)$).
Thus, in this more general setting, all local anomalies
are cancelled by adding to the action the modified \cc\ term
\be
\Gamma^\prime_{WZ} (A,U) = \Gamma^\prime(A^U)-\Gamma^\prime(A),
\label{gammaprime}
\ee
where
\be
\Gamma^\prime(A)= \Gamma(A) + B_d(A_H,A).
\ee
Once gauge invariance is restored, it is
possible to shift the fields in such a way as to decouple $\varphi$,
and give mass to the gauge bosons of $G/H$ (unitary gauge).
The presence of the term~(\ref{gammaprime}) in the generic gauge signals
however the non-renormalizability of the theory. An explicit proof of
non-renormalizability in the 't Hooft-Landau gauge is given in~\cite{Preskill}.
This can also be understood by looking at the diagram giving
rise to the anomaly containing off-shell non-analytic
contributions~\cite{ABJ} that cannot be cancelled by any local counterterm.

The result of these analyses is that, as far as local gauge anomalies
are concerned, effective theories, for which renormalizability is not a requirement,
can be made anomaly free by the addition of an appropriate \cc\ term.
No restriction on the fermion content of the theory is needed, provided that
the first two conditions listed in the introduction are satisfied.

So far, our analysis has been purely local without considering
possible topological obstructions or possible global gauge anomalies.
In fact, a potential problem can arise
in the above construction whenever $\pi_d(G/H)\neq 0$ \cite{CHZ}.
On the other hand, the same condition
of having a non-trivial homotopy group can give rise to global
anomalies~\cite{Wittensu2}. In the next section we will show how
to generalize the above procedure to take into account these global issues
and remove the above topological condition.

\section{Cancellation of Global Gauge Anomalies}

Even a theory free of perturbative anomalies---i.e.\  invariant
under infinitesimal gauge transformations---can still be anomalous under gauge
transformations that are not homotopic to the identity. This was discussed
first for an $SU(2)$ theory in four dimensions in \cite{Wittensu2}
and for other dimensions  in \cite{en} (see also \cite{kiritsis}).
These gauge anomalies are called global, or non-perturbative anomalies.
They can occur whenever $\pi_d(G)\neq 0$.~\footnote{When spacetime is not flat,
or instantons are present, the conditions can become more complicated.
We will not consider these cases.}

Extending the results of \cite{archaeology}, it is always possible
to cancel any global anomaly by coupling the theory to a $G$-valued scalar
field $U(x)$ and adding a suitable \cc\ term $\gamma$ to the action \cite{perca}.
The absence of local anomalies means that $\gamma$, defined as in (\ref{wzdef}),
is zero, independently of $A$, when $g$ is homotopic to a constant.
Every $g$ in a certain homotopy class can be written as $g_1 g'$ where
$g_1$ is a fixed map in that same homotopy class and $g'$ is homotopically trivial.
Then from the cocycle condition (\ref{cocycle}) one sees that
$\gamma(A,g_1 g)=\gamma(A,g_1)$, so $\gamma$ is invariant under continuous
deformations of $g$. In conclusion, $\gamma$ depends only on
the homotopy class  $[g]$ and therefore can be seen as a topological term.~\footnote{The addition
of a topological term to the action is a hallmark of
the existence of $\theta$-sectors, and indeed one can interpret this
anomaly-cancellation mechanism by saying that the anomalous theory
can be quantized by coupling it to a specific $\theta$-sector of the
scalar theory.} From (\ref{cocycle}) we also see that it must define a representation of $\pi_d(G)$:
\be
\gamma(g_1\cdot g_2)=\gamma(g_1)+\gamma(g_2)\,.
\label{additivity}
\ee
As in section II, we are especially interested in the case where the
scalar fields do not have to be introduced \textit{ad hoc}, but are already
present in the theory.
We are thus led to ask: if the $G$ gauge symmetry is spontaneously
broken to $H$ due to a Higgs mechanism, can one cancel the global anomalies
of the low-energy effective theory
by means of a \cc\ term written as a functional of the would-be Goldstone
bosons?
The answer is that this is possible if the fermion representations,
restricted to $H$, are free of global anomalies.

As in the previous section we replace the coset-valued
field $\varphi$ by the $G$-valued field $U$.
Having reformulated the theory in terms of $U$, we can write \cc\ terms
$\gamma(U)$ cancelling any global anomaly, using the method described above.
Of course now we want these terms to depend only on the physical
scalar fields $\varphi$.
This will automatically be the case if the unbroken group $H$ is free
of global anomalies.
Indeed, in this case it follows from (\ref{additivity}) that
$\gamma(U\cdot h)=\gamma(U)+\gamma(h)=\gamma(U)$
and therefore $\gamma$ really only depends on the (homotopy class of the)
coset-valued field $\varphi (x)$.
Global $H$ anomalies will certainly be absent if $\pi_d(H)=0$,
but even if $\pi_d(H)$ is nontrivial the theory may be free of global $H$ anomalies
provided the fermion representations are chosen appropriately.
In this case, global anomalies can be present only when $\pi_d(G/H)\neq 0$.

One may be interested in a general method for calculating the topological
term $\gamma$. Following the discussion in \cite{w} and \cite{en},
it can be written as a \cc\ term, albeit for a larger group.
The construction of the \cc\ term given in (\ref{wzexpl})
demands that the map $U$ be homotopic to a constant.
If $U$ is not homotopic to a constant one can still proceed by embedding
$G$ in a larger group $K$ such that $\pi_d(K)=0$. One then defines a map
$\bar U$ by composing $U$ with this embedding, and a gauge field $\bar A$
by the corresponding embedding of the Lie algebras. For instance,
one can write \cite{WittenGlobal}
\be
\label{matrix}
\bar U = \left( \begin{array}{cc} U & 0\\ 0&1\end{array} \right)
\quad \mbox{and} \quad
\bar A = \left( \begin{array}{cc} A & 0\\ 0&0\end{array} \right) \, .
\ee
The fermion content of $K$ is so chosen  that upon reduction to $G$
gives rise to the required $G$ representations and a number of $G$ singlets (or anomaly-free representations of $G$).

The field $\bar U$ is homotopic to a constant and one can explicitly write the
\cc\ term $\Gamma_{\cc\ }^K(\bar A,\bar U)$ as in (\ref{wzexpl}).
As described above, if the theory is free of perturbative anomalies
we can take simply $\gamma(g)=\Gamma_{\cc\ }^K(\bar A,\bar g)$.
Thus, the \cc\ term canceling the nonperturbative $G$-anomaly can
be calculated
as a perturbative \cc\ term for the larger group $K$, along the lines of~\cite{en}.

Finally, let us consider the generic case when the theory has perturbative anomalies
with only $\delta_i\Gamma(A_H)= 0$, and at the same time $\pi_d(G/H)\neq 0$.
In the last section we have seen, following \cite{Wu,AGG,CHZ}, that it is always possible
to add to the action a local operator $B_d(A_H,A)$
so as to construct a \cc\ term (\ref{gammaprime}) and cancel the perturbative anomalies. This \cc\ term is
well defined when the map $U$ is homotopic to a constant.
We have also described in the previous paragraph a way of writing
a \cc\ term for a larger group $K$ that, when restricted to the
subgroup $G$, makes sense for all maps $U$, irrespective of their homotopy class.
If there are local anomalies, this \cc\ term,  now denoted $\Gamma_{\cc\ }^{\prime K}(\bar A, \bar U)$,
is no longer topological.
In fact,  one can see (e.\ g.\ , by means of (\ref{matrix})) that when $U$
is homotopically trivial, it agrees with $\Gamma^\prime_{WZ}(A,U)$
defined in (\ref{gammaprime}).
There follows that defining
$\Gamma_{\cc\ }^\prime (A,U)=\Gamma_{\cc\ }^{\prime K}(\bar A,\bar U)$
the resulting effective action is defined for all maps $U$,
whether trivial or not, and free of  both local and global gauge anomalies for the
group $G$ spontaneously broken to $H$.~\footnote{Even in the absence of global anomalies,
this construction generalizes that of \cite{CHZ}.}

A physical interpretation of the group $K$ can be given (although not necessary),
along the lines of \cite{hokerfari}. In this case, one can imagine a microscopic
theory with a gauge group $K$ and a completely anomaly free (local and global) fermion spectrum.
The fermions are in representations of $K$ such that, upon the spontaneous breaking of $K$
to $G$, give rise to the required fermions in representations of $G$.
The massive fermions that are integrated out produce the Wess-Zumino term
$\Gamma_{WZ}^{\prime K}(\bar A,\bar U)$ above.

\section{The Standard Model in Six Dimensions}

In a recent paper~\cite{DP}, it was argued that, in the minimal non-supersymmetric
version of the standard model in $d=6$ dimensions, the requirement of
canceling  all gauge anomalies restricts the chiral fermion content
of the theory. In particular, the
cancellation of the $SU(2)$ global anomaly implies a restriction on
the number $N_g$ of matter families:
\bea
N_g \,=\, 0 \, {\mbox{mod}} \, 3\, .
\eea

The field content of a single family is restricted
by the requirement of the cancellation of all irreducible local anomalies---that is, by the requirement of having a vector-like
theory of strong  and gravitational interactions---while the $d=6$ realization of the usual GS mechanism is invokated
in order to eliminate reducible local anomalies. Accordingly, the fields are those in Table~\ref{fields}.

\begin{table}[h]
\begin{center}
\caption{Fermionic field content for each family in~\cite{DP}.
The six-dimensional
chirality is the eigenvalue of $\Gamma_7$.
Here we put just one representative of the allowed
combinations of chiralities; other choices are equivalent as far as the present discussion is concerned.}
\label{fields}
\vspace{0.2cm}
\begin{tabular}{|c|c|c|c|c|}
\hline \hline
 & \ \ Chirality \ \ & $\;\;\;\;U(1)_Y\;\;\;$ & $\;\;\;\;SU(2)_L\;\;$ & $\;\;\;\;SU(3)_c\;\;$ \cr
\hline
$\;\;\;\;Q\;\;\;\;\;$ & $+$ & $1/6$ & 2 & 3 \cr
$\;\;\;\;U\;\;\;\;\;$ & $-$ & $2/3$ & 1 & $3$ \cr
$\;\;\;\;D\;\;\;\;\;$ & $-$ & $-1/3$ & 1 & $3$ \cr
$\;\;\;\;L\;\;\;\;\;$ & $+$ & $-1/2$ & 2 & 1 \cr
$\;\;\;\;E\;\;\;\;\;$ & $-$ & $-1$ & 1 & 1 \cr
$\;\;\;\;N\;\;\;\;\;$ & $-$ & $0$ & 1 & 1 \cr
\hline \hline
\end{tabular}
\end{center}
\end{table}

A detailed inspection of the GS mechanism, performed in~\cite{FPT3},
showed  that the addition of two GS
2-forms---supplemented by the use of the $U(1)_Y$ Goldstone boson
in order to realize the generalized abelian GS of~\cite{DSW}---is sufficient to restore gauge invariance.
After  compactification of the two extra-dimensions, some pseudo-scalar fields, behaving like Peccei-Quinn axions, remain  in the $d=4$ effective field theory as remnants of the GS fields.
Their presence solves the strong $CP$ problem, but imposes a strong bound on the
scale of the GS couplings and therefore  a limit on the
volume of the compact extra-dimensions,
the fundamental scale of which turns out to be in the range of usual GUT theories.

Is it possible
to achieve the cancellation of the reducible anomalies
 with some mechanism which does not leave any axionic remnant in the
low energy $d=4$ effective theory? The machinery discussed in
this paper is suitable for this purpose.

 The choice of fermion content guarantees that the model has no pure gravitational anomaly.  The addition of a local Chern-Simons term shifts mixed gravitational anomalies into those of gauge.
Fermions form a vector-like representation of the group $SU(3)_c$  with no contribution to the anomalies.
We can thus identify the groups of the previous sections with those in~\cite{DP}
\be
\left\{ \begin{array}{lcl}
K&=& SU(4)_L\times U(1)_Y\;,\\
G&=& SU(2)_L\times U(1)_Y\;,\\
H&=& U(1)_{\rm e.m.}\, .
\end{array} \right.
\ee
Since $\pi_6 \left(G/H\right)= \Z_{12}$, we have enlarged $G$ to $K$
for which  $\pi_6 \left(K\right) = 0$.

Each $SU(2)$ doublet in Table~\ref{fields} goes into the fundamental representation of $SU(4)$ (to be decomposed into a $SU(2)$ doublet plus singlets). In addition to the already-present singlets, more singlets are necessary in order to preserve the $U(1)_Y^4$ anomaly.

The subgroup $H$ is not anomaly free, but it is a simple exercise
to verify that, with the field content
in Table~\ref{fields}, fermions form a non-chiral representation of the unbroken group $H$:
 the weaker condition leading to the $B_6$ term is satisfied.
Also all the other conditions stated in the introduction hold, and therefore
the \cc\ term  can be used to cancel all anomalies, as explained in the
previous sections. This term can be explicitly written as
\be
\Gamma^{\prime K}_{WZ} =\,-\,\int_0^1 \di t \,\xi^{\alpha} {\cal A}^{\prime}_{\alpha} [\bar A_t]
 \,,\label{wzdp}
\ee
where $\Gamma^\prime (A) = \Gamma (A)  +  B_6(A_H,A)$, with
\be
B_6(A_0,A_1) =   12\,\int_{\Delta} \di \mu\, \di \lambda\,  \label{bd}
{\rm Str}[A_0,A_1,F_{\mu,\lambda}^2]\,,\nonumber
\ee
and $A_t$ is defined in footnote 3.
Following the notation of~\cite{CHZ}:
\bea
{\rm Str}[C_1,\cdots,C_N]&=& \sum_P \frac{(-1)^{f_P}}{N!} \, \Tr [C_{P_1} \cdots C_{P_N}]\,,\nonumber\\
{\cal A}^{\prime}_{\alpha} (A_t) &=&{\cal A}_{\alpha} (A_t) \,+\,\delta_{\alpha} B_6(A_H,A)\,,\\
F_{\mu,\lambda} &=&  \di A_{\mu,\lambda} + A^2_{\mu,\lambda}\nonumber\,,\\
A_{\mu,\lambda}&=&\mu A_0 + \lambda A_1\nonumber\,.
\eea

In eq.~(\ref{wzdp}), the fields $\xi^{\alpha}$ are the longitudinal components of the
massive gauge bosons $W$ and $Z$. ${\cal A}_{\alpha}$ is the (one-loop) anomaly. All fields
must be thought as $K$-valued, as in~(\ref{matrix}). The integration region $\Delta$ is
a triangle in the $(\mu,\lambda)$ plane with vertices  $(0,1)$, $(1,0)$ and the origin;
$f_P$ is the number of times the permutation $P$ permutes two odd objects.

Thanks to this construction, no GS fields are needed; since there are  no  axions
in the low energy $d=4$ theory, their experimental bounds
do not apply. However, all  global anomalies are
canceled as well, and therefore the interesting prediction on the number of families is lost.
A similar procedure might be relevant also for other six-dimensional extensions discussed in~\cite{Borghini}.

The term $B_6$ in eq.~(\ref{wzdp}) remains, even in the unitary gauge, as a Chern-Simons
coupling between gauge bosons,  and
gives rise to dimension-six operators after compactification to $d=4$.  These operators are
suppressed by a coefficient proportional to the compact volume $R^2$. Even though
they modify, for instance, the photon couplings, we expect their effects to be negligible because
$R \sim $ few TeV$^{-1}$~\cite{AD}.


\acknowledgments

We thank S.~Bertolini, L.~Bonora and M.~Testa for useful discussions. MP also
thanks G.~Tasinato. This work is
 partially supported by  the European TMR Networks HPRN-CT-2000-00148
and HPRN-CT-2000-00152.

\vspace{0.5cm}


\end{document}